\documentclass[aps,prc,preprint,superscriptaddress,showpacs]{revtex4}
\usepackage{multirow}
\usepackage{epsfig}
\usepackage{longtable}
\begin{document}
\title{\boldmath Strangeness magnetic form factor of the proton in the extended chiral quark model}

\author{C. S. An}
\email[]{ancs@ihep.ac.cn} \affiliation{Institute of High Energy Physics and Theoretical Physics 
Center for Science Facilities, CAS, Beijing 100049, China}

\author{B. Saghai}
\email[]{bijan.saghai@cea.fr}
\affiliation{Institut de Recherche sur les lois Fondamentales de l'Univers, 
Irfu/SPhN, CEA/Saclay, F-91191 Gif-sur-Yvette, France}

\date{\today}

\begin{abstract}
{\bf Background:} Unravelling the role played by nonvalence flavors in baryons is crucial in deepening our comprehension
of QCD. 
The strange quark, a component of the higher Fock states in baryons, is an appropriate tool to study 
nonperturbative mechanisms due to the pure sea quark.

{\bf Purpose:} Study the magnitude and the sign of the strangeness magnetic moment $\mu_s$ and the magnetic form factor 
($G_M^s$) of the proton.  

{\bf Methods:} Within an extended chiral constituent quark model, we investigate contributions from all possible five-quark 
components to $\mu_s$ and $G_M^s (Q^2)$ in the four-vector momentum range  $Q^2 \leq 1$ (GeV/c)$^2$. 
The probability of the strangeness component in the proton wave function is calculated employing the $^3 P_0$ model.

{\bf Results:} Predictions are obtained by using input parameters taken from the literature. 
The observables $\mu_s$ and $G_M^s (Q^2)$ are found to be small and negative, consistent with the lattice-QCD findings as well 
as with the latest data released by the PVA4 and HAPPEX Collaborations. 

{\bf Conclusions:} Due to sizeable cancelations among different configurations contributing to the strangeness magnetic 
moment of the proton, it is indispensable to 
{\it i)} take into account all relevant five-quark components and include both diagonal and non-diagonal terms, 
{\it ii)} handle with care the oscillator harmonic parameter $\omega_5$ and the ${s \bar s}$ component probability.
\end{abstract}
%
\maketitle
\section{Introduction}
\label{sec:intro}
Parity-violating electron scattering process, extensively investigated since more than a decade, has been  proven to offer
a unique experimental opportunity in probing the contribution of the strangeness sea to the electromagnetic
properties of the nucleon.
During that period, results from four Collaborations have been released in several publications (for recent reviews
see Refs.~\cite{Armstrong:2012bi,GonzalezJimenez:2011fq}) with the latest ones for each of the Collaborations being: 
SAMPLE (MIT-Bates)~\cite{Spayde:2003nr}, PVA4 (MAMI)~\cite{Baunack:2009gy}, 
G0 (JLab)~\cite{Androic:2009aa}, HAPPEX (JLab)~\cite{Ahmed:2011vp}.
Those experiments allowed extracting linear combinations of electric ($G_E^s$) and magnetic ($G_M^s$) strangeness 
form factors of the proton as a function of four-vector momentum transfer $Q^2$. 

A general trend of the data published before year 2009 was to produce rather small and positive values for $G_M^s (Q^2)$,
especially in the range $(Q^2)\simeq$ 0.1 to 0.5 (GeV/c)$^2$; see e.g. Table I in Ref.~\cite{Young:2006jc}. 
In this latter work a global analysis of World Data of parity-violating electron scattering was performed for 
$Q^2 \lesssim 0.3$ (GeV/c)$^2$ and led to $\mu_s = 0.12 \pm 0.55 \pm 0.07$ nuclear magneton ($\mu_N$).
Another low $Q^2$ global analysis~\cite{Liu:2007yi} disfavored negative $G_M^s$, and still a third one~\cite{Pate:2008va}, 
dedicated to the range $\simeq$ 0.5 to 1.0 (GeV/c)$^2$ produced two sets of solutions with opposite signs.

On the theoretical side, the strangeness contributions to the magnetic moment of the proton have also been intensively
investigated. Few approaches have produced results close to the data, with positive sign, such as
heavy baryon chiral perturbation theory~\cite{Hemmert:1998pi,Hemmert:1999MR}, quenched chiral perturbation 
theory~\cite{Lewis:2002ix}, chiral quark-soliton model~\cite{Silva:2002ej},
Skyrme model~\cite{Xia:2008zzb}, and constituent quark models~\cite{Riska:2005bh,An:2006zf,Kiswandhi:2011ce}.
However, a large number of theoretical results predicted negative values, notably,
meson cloud model~\cite{Forkel:1999kz,Chen:2004rk}, 
chiral quark model~\cite{Hannelius:2000gu,Lyubovitskij:2002ng}, and
unquenched constituent quark model~\cite{Bijker:2012zza}.
A remarkable issue is that the lattice-QCD 
approaches~\cite{Leinweber:1999nf,Leinweber:2004tc,Wang:1900ta,Doi:2009sq,Babich:2010at}
have kept predicting negative strangeness magnetic moment for the proton.
Note that in various works prior to the advent of the first data, the general trend was predicting negative sign for 
the strangeness magnetic moment of the proton $\mu_s$, as reviewed in Refs.~\cite{Beck:2001yx,Beise:2004py}.

In 2009, the PVA4 Collaboration~\cite{Baunack:2009gy}, obtained for the first time a negative sign
value $G_M^s (Q^2=0.22)=-0.14 \pm 0.11 \pm 0.11$; units are (GeV/c)$^2$ for $Q^2$ and nuclear magnetons for $G_M^s$.
More recently the HAPPEX Collaboration~\cite{Ahmed:2011vp} reported a small but also negative sign at 
higher $Q^2$, namely, $G_M^s (Q^2=0.624)=-0.070 \pm 0.067\mu_N$.

The present work is motivated by interpreting the recent data~\cite{Baunack:2009gy,Ahmed:2011vp} on
$G_M^s (Q^2)$ within an extended chiral constituent quark model ($E \chi CQM$). 

Our starting point was the idea put forward by Zou and Riska~\cite{Zou:2005xy} according to which 
the strangeness magnetic moment of the proton could be explained by including five-quark Fock 
components in the proton wave function. 
They showed that a positive strangeness magnetic moment of the proton can rise from the $\bar s$ 
being in the ground state and the four-quark subsystem $uuds$ in the $P$-state, while 
$\bar s$ in the $P$-state and the four-quarks in their ground state would lead to a negative
value for $\mu_s$. 
Then that approach was developed and extended to the strangeness contributions to spin of the
proton~\cite{An:2005cj}, magnetic moments of baryons~\cite{An:2006zf}, electromagnetic and strong decays
of baryon resonances~\cite{Li:2005jn,Li:2006nm,An:2010wb,An:2011sb}.
The main outcome of those studies is that the higher Fock components play important roles in describing 
the properties of baryons and their resonances.

However, in Ref.~\cite{Zou:2005xy} only contributions from the diagonal matrix elements 
$\langle uuds\bar{s}|\hat{\mu}_{s}|uuds\bar{s}\rangle$ were included,
while the non-diagonal transition between three-quark and strangeness components of the proton 
$\langle uud|\hat{\mu}_{s}|uuds\bar{s}\rangle$ also contributes. 
In fact, the diagonal contributions are proportional to the probability of corresponding strangeness 
component $P_{s\bar{s}}\equiv A_{s\bar{s}}^{2}$, but the non-diagonal contributions are proportional to 
the product of probability amplitudes of three- and five-quark components $A_{3q}A_{s\bar{s}}$. 
Generally, the latter is more significant than the former, given that the proton is mainly composed of 
three-quark component.
In Ref.~\cite{An:2006zf}, the non-diagonal contributions were taken into account, but on the one
hand, only the lowest strangeness component, with the four-quark subsystem in the P-state was considered,
and on the other hand, the probability amplitudes for strangeness components in the proton were treated as free 
parameters in order to obtain a positive value for $\mu_s$.

In the present work, the probability amplitudes, a crucial ingredient in the extended chiral constituent quark model,
are calculated within the most commonly accepted $q\bar{q}$ pair creation mechanism, namely, the $^{3}P_{0}$ 
model. Then, the $q\bar{q}$ pair is created anywhere in space with the quantum numbers of the QCD vacuum 
$0^{++}$, corresponding to $^{3}P_{0}$~\cite{Le Yaouanc:1972ae}. 
This model has been successfully applied to the decay of mesons and baryons~\cite{Le Yaouanc:1973xz,Kokoski:1985is}, 
and has recently been employed to analyze the sea flavor content of the ground states of the $SU(3)$ octet
baryons~\cite{An:2012kj}. 
Note that in the $SU(3)$ symmetric case, the ratio of probabilities for five-quark components with strange and light 
quark-antiquark pairs is $1/2$~\cite{An:2006zf}, while by taking into account the $SU(3)$ symmetry breaking effects, 
we determined~\cite{An:2012kj} that ratio to be  
$P_{s\bar{s}}/(P_{u\bar{u}}+P_{d\bar{d}})=0.057/(0.098+0.216)\sim 0.18$ and putting $P_{s\bar{s}} \sim 6\%$.

Moreover, we calculate both diagonal and non-diagonal terms for all relevant five-quark configurations and
removed contributions from the center-of-mass motion of the quark clusters, as emphasized 
recently~\cite{Kiswandhi:2011ce}.

Finally, we underline that all of the input parameters are taken consistently from the literature.

The present paper is organized as follows. 
In Section \ref{sec:frame}, we present our theoretical framework, which includes the wave
function and the strangeness magnetic moment of the proton within our extended constituent quark model.
Our numerical results for the strangeness magnetic moment and form factor of the proton are reported in 
Section \ref{sec:result}, where we give the input parameters, discuss the role of various ingredients of our 
approach and proceed to comparisons with findings by other authors. 
Finally, Section \ref{sec:end} contains summary and conclusions.
%
%
\section{Theoretical framework}
\label{sec:frame}
In this section, we first briefly review the method to derive the wave function of the proton in the extended chiral 
constituent quark model (Sec~\ref{sec:wfn}), and then present the formalism for the strangeness magnetic moment of 
the proton (Sec~\ref{sec:smm}).
\subsection{Wave function of the proton}
\label{sec:wfn}
In our extended chiral constituent quark model, the wave function of the proton can be expressed as
\begin{equation}
 |\psi\rangle_{p}=\frac{1}{\mathcal{\sqrt{N}}}\left(|3q\rangle+
 \sum_{i,n_{r},l}C_{in_{r}l}|5q,i,n_{r},l\rangle \right)\,.
\label{wfn}
\end{equation}
The first term in Eq.~(\ref{wfn}) is just the conventional wave function for the proton with three light constituent quarks, 
which reads
\begin{eqnarray}
|3q\rangle=\frac{1}{\sqrt{2}}[1^{3}]_{C}\phi_{000}(\vec{\xi}_{1})
\phi_{000}(\vec{\xi}_{2})(\varphi_{\lambda}^{p}\chi_{\lambda}+
\varphi_{\rho}^{p}\chi_{\rho})\,,
\end{eqnarray}
where $[1^{3}]_{C}$ denotes the $SU(3)$ color singlet, $\varphi_{\lambda(\rho)}^{p}$ the mixed symmetric flavor wave 
functions of the proton, and $\chi_{\lambda(\rho)}$ the mixed symmetric spin wave functions for configuration
$[21]_{S}$ with spin $1/2$ for a three-quark system. 
And $\phi_{000}(\vec{\xi}_{i})$ are the orbital wave functions with the quantum numbers $n_{r},l,m$ denoted by corresponding 
subscripts; $\vec{\xi}_{i}$ are the Jacobi coordinates defined by
\begin{equation}
\vec{\xi}_{1}=\frac{1}{\sqrt{2}}(\vec{r}_{1}-\vec{r}_{2});~~ 
\vec{\xi}_{2}=\frac{1}{\sqrt{6}}(\vec{r}_{1}+\vec{r}_{2}-2\vec{r}_{3})\,.
\label{xi}
\end{equation}
The second term in Eq.~(\ref{wfn}) is a sum over all possible five-quark Fock components with $q\bar{q}$ pairs; 
$q \equiv u,d,s$.
$n_{r}$ and $l$ denote the inner radial and orbital quantum numbers, respectively. 
As discussed in Ref.~\cite{An:2012kj}, here we only consider the case for $n_{r}=0$ and $l=1$, since probabilities
of higher radial excitations in the proton should be very small, and those of higher orbital excitations vanish.
Different possible orbital-flavor-spin-color configurations of the four-quark subsystems in the five-quark system 
with $n_{r}=0$ and $l=1$ are numbered by $i$; $i=1,\cdots,17$.
Finally, $C_{in_{r}l}/\sqrt{\mathcal{N}}\equiv A_{in_{r}l}$ represents the probability amplitude for the corresponding 
five-quark component, which can be calculated by
\begin{equation}
C_{in_{r}l}=\frac{\langle QQQ(Q\bar{Q}),i,n_{r},l|\hat{T}|QQQ\rangle}{M_{p}-E_{in_{r}l}}\,,
\end{equation}
where
\begin{eqnarray}
\mathcal{N} & \equiv & 1+ \sum_{i=1}^{17} \mathcal{N}_i  
= 1+ \sum_{i=1}^{17}C_{in_{r}l}^{2},
\label{norm}
\end{eqnarray}
and $\hat{T}$ is a transition coupling operator of the $^{3}P_{0}$ model
\begin{eqnarray}
 \hat{T}&=&-\gamma\sum_{j}\mathcal{F}_{j,5}^{00}\mathcal{C}_{j,5}^{00}\mathcal{C}_{OFSC}\sum_{m}
\langle1,m;1,-m|00\rangle\nonumber\\
&&\chi^{1,m}_{j,5}
\mathcal{Y}^{1,-m}_{j,5}(\vec{p}_{j}-\vec{p}_{5})b^{\dag}(\vec{p}_{j})d^{\dag}(\vec{p}_{5})\,,
\label{op}
\end{eqnarray}
with $M_{p}$ the physical mass of the proton. 

Wave functions of the five-quark components can be classified into two categories by four-quark subsystems
being in their S-state
\begin{eqnarray}
|5q,i,0,1\rangle&=&
\sum_{abc}\sum_{s_{z}mm's'_{z}}C^{\frac{1}{2}
\frac{1}{2}}_{1s_{z},jm}C^{jm}_{1m',\frac{1}{2}s'_{z}}C^{[1^4]}_{[31]_{a}
[211]_{a}}\nonumber\\
&&C^{[31]_{a}}_{[F]_{b}[S]_{c}}[F]_{b}[S]_{c}[211]_{C,a}
\bar Y_{1m'} \bar \chi_{s_{z'}}\nonumber\\
&&\Phi(\{\vec \xi_i\})\, ,
\label{wfc1}
\end{eqnarray}
and P-state
\begin{eqnarray}
|5q,i,0,1\rangle&=&
\sum_{abcde}\sum_{Ms'_{z}ms_{z}}C^{\frac{1}{2}\frac{1}{2}}
_{JM,\frac{1}{2}s'_{z}}C^{JM}_{1m,Ss_{z}}C^{[1^{4}]}_{[31]_{a}
[211]_{a}}
\nonumber\\
&&C^{[31]_{a}}_{[31]_{b}[FS]_{c}}C^{[FS]_{c}}_{[F]_{d}[S]_{e}}
[31]_{X,m}(b)
[F]_{d}\nonumber\\
&&[S]_{s_{z}}(e)[211]_{C}(a)\bar\chi_{s'_{z}}
\Phi(\{\vec \xi_i\})
\, ,
\label{wfc2}
\end{eqnarray}
where the flavor, spin, color and orbital wave functions of the four-quark subsystem are denoted by the Young patterns. 
The coefficients $C^{\frac{1}{2}\frac{1}{2}}_{1s_{z},jm}$ and $C^{jm}_{1m',\frac{1}{2}s'_{z}}$ in Eq.~(\ref{wfc1}), 
and $C^{\frac{1}{2}\frac{1}{2}}_{JM,\frac{1}{2}s'_{z}}$ and $C^{JM}_{1m,Ss_{z}}$ in Eq.~(\ref{wfc2}) are Clebsch-Gordan
coefficients for the angular momentum, and others are Clebsch-Gordan coefficients of $S_{4}$ permutation group. 
$\bar Y_{1m'}$ and $\bar \chi_{s_{z'}}$ represent the wave functions of the antiquark.
$\vec{\xi}_{i}$ denote the Jacobi coordinates for a five-quark system, analogous to the ones in Eq.~(\ref{xi}), and
$\vec{\xi}_{i}$ are defined as
\begin{eqnarray}
 \vec{\xi_{i}}&=&\frac{1}{\sqrt{i+i^{2}}}\left
 (\sum_{j=1}^{i}\vec{r}_{j}-i\vec{r}_{i}\right),~i=1,\cdots,4\,.
 \label{jac}
\end{eqnarray}

Finally, the energies of five-quark components with quantum numbers $n_{r}=0$ and $l=1$ in constituent quark model
can be expressed as
\begin{equation}
E_{i,0,1}=E_{0}+\delta_{m}^{i}+\langle H_{hyp}\rangle_{i}\,, 
\end{equation}
where $E_{0}$ is a commonly shared energy of the $17$ different five-quark configurations, $\delta_{m}^{i}$ the energy 
deviation caused by the $s\bar{s}$ pairs, and $\langle H_{hyp}\rangle_{i}$ denote matrix elements of the quarks hyperfine 
interactions in the five-quark configurations. In this work, we employ the 
hyperfine interactions mediated by Goldstone-boson exchange~\cite{Glozman:1995fu},
\begin{eqnarray}
 H_{h}&=&-\sum_{i<j}\vec{\sigma}_{i}\cdot\vec{\sigma}_{j}
                    \Big [ \sum_{a=1}^{3}V_{\pi}(r_{ij})\lambda^{a}_{i}\lambda^{a}_{j}+
                   \sum_{a=4}^{7}V_{K}(r_{ij})\lambda^{a}_{i}\lambda^{a}_{j}\nonumber\\
                   &&+V_{\eta}(r_{ij})\lambda^{8}_{i}\lambda^{8}_{j} \Big ]\,,
\label{hyp}
\end{eqnarray}
where $V_{M}$ are the corresponding strength of the $M$($\equiv \pi,~K,~\eta $) meson-exchange interactions, and 
$\lambda^{a}_{i(j)}$ the Gell-Mann matrices in $SU(3)$ color space.

Explicit matrix elements of $\hat{T}$ and the energies $E_{i,0,1}$ were derived in Ref.~\cite{An:2012kj}, 
here we employ those results for calculations of the probability amplitudes of the strangeness components in 
the proton.
%
%
\subsection{Strangeness magnetic moment of the proton}
\label{sec:smm}
In our model, calculations of the strangeness magnetic moment of the proton can be divided into two parts, namely, 
the diagonal and non-diagonal contributions. 
The former can be defined as the matrix elements of the following operator in the strangeness
components of the proton
\begin{equation}
 \hat{\mu}_{s}^{D}=\frac{M_{p}}{m_{s}}\sum_{i}\hat{S}_{i}
 \left(\hat{l}_{iz}+\hat{\sigma}_{iz}\right)\,,
 \label{mud}
\end{equation}
where $\hat{S}_{i}$ is an operator acting on the flavor space, with the eigenvalue $+1$ for a strange quark, $-1$ 
for an anti-strange quark, and $0$ for the light quarks. 
Note that the operator $\hat{\mu}_{s}^{D}$ is in unit of the nuclear magneton.

The non-diagonal contributions of the strangeness magnetic moment, which involve $s\bar{s}$ pair annihilations and
creations, are obtained as matrix elements of the operator
\begin{equation}
 \hat{\mu}_{s}^{ND}=2M_{p}\sum_{i}\frac{\hat{S}_{i}}{2}\mathcal{C}_{OFSC}
 \vec{r}_{i}\times\hat{\sigma_{i}}\,,
 \label{mund}
\end{equation}
where $\hat{\mu}_{s}^{ND}$ is also in unit of the nuclear magneton. 
$\mathcal{C}_{OFSC}$ is an operator to calculate the overlap between the orbital, flavor, spin and color wave 
functions of the residual three-quark in the five-quark components after $s\bar{s}$ annihilation and the three-quark 
component of the proton.

%

\begin{table}[hbt]
\caption{\footnotesize Diagonal ($\mu^{D}_{s}$) and non-diagonal ($\mu^{ND}_{s}$)contributions of different five-quark 
configurations to the strangeness magnetic moment of the proton. 
Notice that the full expressions are obtained by multiplying each term by 
$\frac{M_{p}}{m_{s}}P_{s\bar{s}}^{i}$ for $\mu_{s}^{D}$ and
by $\frac{4M_{p}}{9\omega_{5}}15^{3/4}C_{35}A_{3q}A_{s\bar{s}}^{i}$ for $\mu_{s}^{ND}$.
The last column gives the flavor-spin overlap factors.
\label{smm}}
\scriptsize
\vspace{0.5cm}
\begin{tabular}{llccc} 
\hline\hline
Category & Configurations & $\mu^{D}_{s}$ & $\mu^{ND}_{s}$ & $C_{FS}^{i}$ \\
\hline
 i) $[31]_{X}[22]_{S}$: & & &  &  \\
 & $[31]_{X}[4]_{FS}[22]_{F}[22]_{S}$      & $1/2$    &  $2\sqrt{3}/3$  & $\sqrt{2}/4$ \\
 & $[31]_{X}[31]_{FS}[211]_{F}[22]_{S}$    & $13/24$  &  $2\sqrt{3}/3$  & $\sqrt{2}/4$ \\
 & $[31]_{X}[31]_{FS}[31]_{F}[22]_{S}$     & $13/24$  &  $2\sqrt{3}/3$  & $\sqrt{2}/4$ \\
%
%
 ii) $[31]_{X}[31]_{S}$: & & &  &  \\
 & $[31]_{X}[4]_{FS}[31]_{F}[31]_{S}$      & $-1$     &  $-2\sqrt{3}/3$ & $1/2$ \\
 & $[31]_{X}[31]_{FS}[211]_{F}[31]_{S}$    & $-1$     &  $-2\sqrt{3}/3$ & $1/2$ \\
 & $[31]_{X}[31]_{FS}[22]_{F}[31]_{S}$     & $-1$     &  $-2\sqrt{3}/3$ & $1/\sqrt{6}$ \\
 & $[31]_{X}[31]_{FS}[31]_{F}[31]_{S}$     & $-1$     &  $2/3$          & $-\sqrt{3}/6$ \\
%
%
 iii) $[4]_{X}[22]_{S}$: & & &  &  \\
 & $[4]_{X}[31]_{FS}[211]_{F}[22]_{S}$     & $-1/6$   &  $\sqrt{6/5}$   & $\sqrt{3}/4$  \\
 & $[4]_{X}[31]_{FS}[31]_{F}[22]_{S}$      & $-1/6$   &  $\sqrt{6/5}$   & $\sqrt{3}/4$ \\
%
%
 iv) $[4]_{X}[31]_{S}$: & & &  &  \\
 & $[4]_{X}[31]_{FS}[211]_{F}[31]_{S}$     & $-1$     &  $-\sqrt{6/5}$  & $\sqrt{6}/4$  \\

 & $[4]_{X}[31]_{FS}[22]_{F}[31]_{S}$      & $-1$     &  $-2/\sqrt{5}$  & $1/2$ \\

 & $[4]_{X}[31]_{FS}[31]_{F}[31]_{S}$      & $-1$     &  $2/\sqrt{10}$  & $-\sqrt{2}/4$ \\

\hline 
\hline
\end{tabular}
\end{table}

As reported previously~\cite{An:2012kj}, among the seventeen possible different five-quark configurations, 
the probability amplitudes of twelve of them with $s\bar{s}$ pairs are nonzero in the proton.
Those configurations can be classified in four categories (Table~\ref{smm}) with respect the orbital and spin 
wave functions of the four-quark subsystem, namely, configurations with: 
{\it i)} $[31]_{X}$ and $[22]_{S}$; 
{\it ii)} $[31]_{X}$ and $[31]_{S}$; 
{\it iii)} $[4]_{X}$ and $[22]_{S}$; {\it iv)} $[4]_{X}$ and $[31]_{S}$. 
Contributions of these four different kinds of configurations are described below.

{ $\bf i)~[31]_{X}$ {\bf and} $\bf [22]_{S}$:}
The total spin of the four-quark subsystem is $0$, therefore the diagonal matrix elements $\langle\mu_{s}^{D}\rangle$ are only 
from contributions due to the four-quark orbital angular momentum and spin of the antiquark, the resulting matrix elements are
\begin{equation}
 \langle\mu_{s}^{D}\rangle_{i}=\frac{M_{p}}{3m_{s}}\Big [1+
 2\langle\sum_{j=1}^{4}\hat{l}_{jz}\hat{S}_{j}\rangle_{i}\Big ]P_{s\bar{s}}^{i}\,,
\end{equation}
where $P_{s\bar{s}}^{i}$ is the probability of the $i^{th}$ strangeness component in the proton.
And for the non-diagonal matrix element $\langle\mu_{s}^{ND}\rangle$, explicit calculations lead to 
\begin{equation}
 \langle\mu_{s}^{ND}\rangle_{i}=\frac{15^{3/4}M_{p}C_{35}}
 {27\omega_{5}}16\sqrt{6}\mathcal{C}_{FS}^{i}A_{3q}A_{s\bar{s}}^{i}\,,
\end{equation}
where $A_{3q}$ and $A_{s\bar{s}}^{i}$ denote the probability amplitudes of the three-quark and the $i^{th}$ strangeness 
components in the proton, and $\mathcal{C}_{FS}^{i}$ is the corresponding flavor-spin overlap factor for the $i^{th}$
strangeness component.
$C_{35}$, common to all different strangeness components, is the overlap between the orbital wave function of the residual 
three-quark in the strangeness component after $s\bar{s}$ annihilation and that of the three-quark component, and reads
\begin{equation}
 C_{35}=\left(\frac{2\omega_{3}\omega_{5}}{\omega_{3}^{2}+\omega_{5}^{2}}\right)^{3}\,,
\end{equation}
with $\omega_{3}$ and $\omega_{5}$ the harmonic oscillator parameters of three- and five-quark components. 
Note that the expression for $C_{35}$ above differs by a factor of 
$[{2\omega_{3}\omega_{5}}/{(\omega_{3}^{2}+\omega_{5}^{2})}]^{3/2}$ 
from that introduced in, e.g. Refs.~\cite{Riska:2005bh,An:2006zf}, due to the proper handling of the center-of-motion 
in the present work. 

{$\bf ii)~[31]_{X}$ {\bf and} $\bf [31]_{S}$:}
The total spin of the four-quark subsystem is $1$, combined to the orbital angular momentum $L_{[31]_{X}}=1$, the total angular
momentum of the four-quark subsystem can be $J=0,1,2$, and to form the proton spin $1/2$, only
the former two are possible alternatives.
In the present case, we take the lowest one $J=0$. Accordingly, the four-quark  subsystem cannot contribute to 
$\mu_{s}$, and the resulting matrix elements are
\begin{eqnarray}
 \langle{\mu}_{s}^{D}\rangle_{i}&=&-\frac{M_{p}}{m_{s}}P_{s\bar{s}}^{i}\,, \\
 \langle\mu_{s}^{ND}\rangle_{i}&=&-\frac{15^{3/4}M_{p}C_{35}}
 {27\omega_{5}}16\sqrt{3}\mathcal{C}_{FS}A_{3q}A_{s\bar{s}}^{i}\,,
\end{eqnarray}

{$\bf iii)~[4]_{X}$ {\bf and} $\bf [22]_{S}$:}
 Given that the total angular momentum of the four-quark subsystem is $0$, it does not contribute
to $\mu_{s}$. 
Consequently, once we remove the contributions of the momentum of the proton center-of-mass motion,
we obtain the following matrix elements:
\begin{eqnarray}
 \langle{\mu}^{D}_{s}\rangle_{i}&=&-\frac{M_{p}}{m_{s}}\Big (\frac{1}{5}-\frac{2}{15}
 \langle\sum_{j=1}^{4}\hat{S}_{j}\rangle_{i}\Big )P_{s\bar{s}}^{i}\,, \\
 \langle\mu_{s}^{ND}\rangle_{i}&=&\frac{15^{3/4}M_{p}C_{35}}
 {9\omega_{5}}16\sqrt{\frac{2}{5}}\mathcal{C}_{FS}^{i}A_{3q}A_{s\bar{s}}^{i}\,.
\end{eqnarray}

{$\bf iv)~[4]_{X}$ {\bf and} $\bf [31]_{S}$:}
The total spin of the four-quark subsystem should be $S_{[31]}=1$, here we assume that the combination of
$S_{[31]}$ with orbital angular momentum of the antiquark leads to
$J=S_{4}\oplus L_{\bar{q}}=0$, then matrix elements read 
\begin{eqnarray}
 \langle\hat{\mu}_{s}^{D}\rangle_{i}&=&-\frac{M_{p}}{m_{s}}P_{s\bar{s}}^{i}\,, \\
 \langle\mu_{s}^{ND}\rangle_{i}&=&-\frac{15^{3/4}M_{p}C_{35}}
 {9\omega_{5}}16\sqrt{\frac{1}{5}}\mathcal{C}_{FS}^{i}A_{3q}A_{s\bar{s}}^{i}\,.
\end{eqnarray}

Accordingly, explicit calculations of the matrix elements 
$\langle\sum_{j=1}^{4}\hat{l}_{jz}\hat{S}_{j}\rangle_{i}$,
$\langle\sum_{j=1}^{4}\hat{S}_{j}\rangle_{i}$, and $\mathcal{C}_{FS}^{i}$
lead to the results shown in Table~\ref{smm}.
%
%
\section{Numerical results and discussion}
\label{sec:result}
As already mentioned, numerical results reported here were obtained using 
input parameters (Table~\ref{fp}) taken from the literature, as commented below.

\begin{table*}[htb]
\caption{\footnotesize Input parameters (in MeV).
\label{fp}}
\scriptsize
\vspace{0.5cm}
\begin{tabular}{ccc} 
\hline\hline
  Parameter  & value           & Ref.  \\
\hline
  $m_s$      &   $460$         & \cite{Glozman:1995fu} \\
  $\delta m$ &   $120$         & \cite{Glozman:1995fu} \\
   $E_0$     &   $2 127$       & \cite{An:2011sb} \\
   $V$       &   $570 \pm 46$  & \cite{An:2011sb} \\
  $A_0$      &   $29$          & \cite{Glozman:1995fu} \\
  $B_0$      &   $20$          & \cite{Glozman:1995fu} \\
  $C_0$      &   $14$          & \cite{Glozman:1995fu} \\
  $A_1$      &  $45$           & \cite{Glozman:1995fu} \\
  $B_1$      &  $30$           & \cite{Glozman:1995fu} \\
  $C_1$      &  $20$           & \cite{Glozman:1995fu} \\
  $\omega_3$ &   $246~ \& ~340 $ & \cite{An:2010wb,An:2011sb,An:2013zoa} \\
  $\omega_5$ &   $225~ \& ~600$  & \cite{An:2010wb,An:2011sb,An:2013zoa} \\
\hline 
\hline
\end{tabular}
\end{table*}

For the mass of the strange quark $m_s$ and the mass difference between constituent strange and light 
quarks $\delta m=m_s-m$, we adopted the commonly used values~\cite{Glozman:1995fu}. 
The energy shared by five-configurations between quarks $E_0$, in the absence of hyperfine interaction,
and the term due to the transition between three- and five-quark components ($V$) are taken from our
previous work~\cite{An:2012kj}, which allowed reproducing the experimental data for the proton flavor 
asymmetry $\bar{d} - \bar{u}$. 
The matrix elements of the flavor operators, are linear combinations of the spatial matrix elements,
$A_i$, $B_i$ and $C_i$, $i$=0,1 ; the numerical values of which were fixed to those determined in 
Ref.~\cite{Glozman:1995fu}.

The last two parameters in Table~\ref{fp} are the harmonic oscillator parameters, $\omega_{3}$ and $\omega_{5}$, for the 
three- and five-quark components, respectively, in baryons. 
The parameter $\omega_{3}$ can be inferred from the empirical radius of the proton {\it via}
$\omega_{3}=1/\sqrt{\langle r^{2}\rangle}$, which yields $\omega_{3}\simeq 246$~MeV for $\sqrt{\langle r^{2}\rangle}=1$~fm.
However, the value of $\omega_{5}$ is rather difficult to determine empirically. 
As discussed in Ref.~\cite{An:2013zoa}, the ratio 
\begin{equation}
 R=\frac{\omega_{5}} {\omega_{3}}\,,
\end{equation}
can be larger or smaller than $1$. 
Consequently, we used two sets for $R$ to get the numerical results, 
\begin{itemize}
  \item {\bf Set~I:} $\omega_{3}=246$~MeV and $R=\sqrt{5/6} \simeq 0.91$ from setting the confinement strength of three- and 
  five-quark configurations to be the same value~\cite{An:2013zoa}, leading to $\omega_{5} \simeq 225$~MeV and 
  $C_{35} \simeq 0.99$.
  \item {\bf Set~II:} $\omega_{3}=340$~MeV and $\omega_{5}=600$~MeV, values adopted to reproduce the data
for electromagnetic and strong decays of several baryon resonances~\cite{An:2010wb,An:2011sb}, corresponding to $R \simeq 1.76$
and $C_{35} \simeq 0.63$. 
\end{itemize}

Finally, a crucial ingredient of our approach is the probability of the strange quark-antiquark components $P_{s\bar{s}}$, 
which is often left as free parameter. 
Here, we calculated it within the $^3P_0$ formalism~\cite{Le Yaouanc:1972ae,Le Yaouanc:1973xz,Kokoski:1985is}.
Then, that probability turns out~\cite{An:2011sb} to be $P_{s\bar{s}}=5.7 \pm 0.6 \%$, for $V=570 \pm 46$ MeV.

In the following two sections we report our results for the strangeness magnetic moment $\mu_s$ and 
magnetic form factor $G_M^s$ of the proton and compare them with the latest data and few most recent / relevant 
theoretical investigations. 

%
\subsection{Strangeness magnetic moment of the proton}
\label{sec:mmp}
Our results for diagonal and non-diagonal components of $\mu_s$ are reported in Table~\ref{nsmm}, for the central value 
$V=570$~MeV and the two Sets with respect to the [$R$, $\omega_{3}$, $\omega_{5}$] ensembles presented above.

%
\begin{table*}[ht]
\caption{\footnotesize Diagonal $\mu^{D}_{s}$ and non-diagonal $\mu^{ND}_{s}$ contributions to the strangeness magnetic 
moment of the proton from each configuration for Sets I and II, with $A^i_{s\bar{s}}$ the probability amplitude and 
$P^i_{s\bar{s}}/P^{tot}_{s\bar{s}}$ the relative weight of the strangeness probability in the proton; 
$P^{tot}_{s\bar{s}}=\sum_{i=1}^{12} P^i_{s\bar{s}}$.
\label{nsmm}}
\scriptsize
\vspace{0.5cm}
\begin{tabular}{llccccc} 
\hline\hline

             & & & &     & Set~I  & Set~II \\
\hline

Category & Configuration & $A^i_{s\bar{s}}$ & $P^i_{s\bar{s}}/P^{tot}_{s\bar{s}}$ &$\mu^{D}_{s}$& $\mu^{ND}_{s}$  &  
             $\mu_{s}^{ND}$ \\
         &                &                   & $(\%)$ & ($\mu_N$) & ($\mu_N$)  &  ($\mu_N$) \\
\hline 
 i) $[31]_{X}[22]_{S}$: & & &  &  \\
 & $[31]_{X}[4]_{FS}[22]_{F}[22]_{S}$  & $-0.099$   & $17$       & ~~$0.0100$      &  $-1.0043$      &  $-0.2403$ \\
 & $[31]_{X}[31]_{FS}[211]_{F}[22]_{S}$ & $-0.060$  & $~6$       & ~~$0.0040$      &  $-0.6121$      &  $-0.1464$ \\
 & $[31]_{X}[31]_{FS}[31]_{F}[22]_{S}$  & $-0.051$  & $~5$       & ~~$0.0029$      &  $-0.5196$      &  $-0.1243$ \\
 ~~~~{\it Subtotal 1} &                 &         & ${\it 28}$ & ${\it ~~0.0169}$& ${\it -2.1360}$ &  ${\it -0.5110}$ \\
%
%
 ii) $[31]_{X}[31]_{S}$: & & &  &  \\
 & $[31]_{X}[4]_{FS}[31]_{F}[31]_{S}$  &  $-0.079$  & $11$  & $-0.0128$     &  ~$0.8033$    &  ~$0.1922$ \\
 & $[31]_{X}[31]_{FS}[211]_{F}[31]_{S}$ & $-0.057$  & $~6$  & $-0.0066$     &  ~$0.5767$    &  ~$0.1380$ \\
 & $[31]_{X}[31]_{FS}[22]_{F}[31]_{S}$  & $-0.042$  & $~3$  & $-0.0036$     &  ~$0.4273$    &  ~$0.1022$ \\
 & $[31]_{X}[31]_{FS}[31]_{F}[31]_{S}$  & $~~0.028$ & $~1$  & $-0.0016$     &  ~$0.1617$    &  ~$0.0387$ \\
 ~~~~{\it Subtotal 2} &                 &     & ${\it 21}$ & ${\it -0.0246}$& ${\it ~~1.9690}$ &  ${\it ~~0.4711}$ \\
%
%
 iii) $[4]_{X}[22]_{S}$: & & &  &  \\
 & $[4]_{X}[31]_{FS}[211]_{F}[22]_{S}$  & $~0.092$   & $15$  & $-0.0029$     &  ~~$0.8894$    &  ~~$0.2128$ \\
 & $[4]_{X}[31]_{FS}[31]_{F}[22]_{S}$   &  $~0.081$  & $11$  & $-0.0022$     &  ~~$0.7772$    &  ~~$0.1859$ \\
 ~~~~{\it Subtotal 3} &                 &         & ${\it 26}$ & ${\it -0.0051}$& ~~${\it 1.6666}$ &  ~~${\it 0.3987}$ \\
%
%
 iv) $[4]_{X}[31]_{S}$: & & &  &  \\
 & $[4]_{X}[31]_{FS}[211]_{F}[31]_{S}$  & $~0.088$   & $13$  & $-0.0157$     &  $-0.8450$   &  $-0.2022$ \\

 & $[4]_{X}[31]_{FS}[22]_{F}[31]_{S}$   & $~0.066$   & $~8$  & $-0.0089$     &  $-0.5202$   &  $-0.1244$\\

 & $[4]_{X}[31]_{FS}[31]_{F}[31]_{S}$  &  $-0.044$  & $~3$  & $-0.0039$     &  $-0.2426$   &  $-0.0580$ \\
 ~~~~{\it Subtotal 4} &                 &         & ${\it 24}$ & ${\it -0.0285}$& ${\it -1.6078}$ &  ${\it -0.3846}$ \\  \\            
 & TOTAL                               &     -      & $100$  & $-0.0413$     &  $-0.1082$  &  $-0.0258$\\

\hline 
\hline
\end{tabular}
\end{table*}

In Table~\ref{nsmm} the first column shows the four categories and the second one the associated configurations. 
Accordingly, contributions from each one of the twelve configurations are reported.
Probability amplitudes, calculates within the $^3P_0$ model are depicted in the third column.
The fourth column gives the relative weight for each configuration in $P_{s\bar{s}}=5.7\%$.
The diagonal terms (fifth column), not depending on $\omega_{5}$, are identical for the two Sets.
Finally, the last two columns correspond to the contributions from non-diagonal terms for Sets I and II, respectively.
Several features deserve comments, which will also be useful in shedding light on the results from other sources.

\begin{itemize}
  \item {$\bf A^i_{s\bar{s}}$:} The probability amplitudes for all $[31]_{X}$ configurations are negative, 
  except for the one with flavor-spin wave function $[31]_{FS}[31]_{F}[31]_{S}$, while those for configurations
  with $[4]_{X}$ are positive, except for the $[31]_{FS}[31]_{F}[31]_{S}$ configuration.
  \item {$\bf P^i_{s\bar{s}}/P^{tot}_{s\bar{s}}$:} The total contribution of each category is around $24 \pm 4\%$, 
  so comparable to each other. 
  However, the probabilities of individual configurations span from $1\%$ to $17\%$.
  \item {$\bf \mu^{D}_{s}$:} The diagonal terms are positive in the first category and negative in the other three.
  The absolute values from one configuration to another show variations reaching almost one order of magnitude.
  \item {$\bf \mu_{s}^{ND}$:} The difference between Sets I and II per configuration is merely due to the different
  [$\omega_{3}$, $\omega_{5}$] ensembles used in the present work. 
  Non-diagonal terms have opposite signs with respect to the corresponding diagonal ones in all categories, except the 
  last one. 
  Per configuration, the magnitude of non-diagonal term is larger, in some cases by two orders of magnitudes, 
  than that of the corresponding diagonal term. 
  \item {$\bf \mu^{D}_{s} + \mu_{s}^{ND}$:}
  Accordingly, the sum of the diagonal and non-diagonal terms per configuration is dominated by far by the 
  non-diagonal term.
  However, it is important to underline the following point: the last line in Table~\ref{nsmm} shows that, due to 
  significant cancelations among the non-diagonal terms from various configurations, the ratio of the sum of 
  non-diagonal terms (-0.1082 and -0.0258) over that of the diagonal ones (-0.0413), is 2.6 (Set~I) or 0.6 (Set~II),
  so very significantly different from that ratio per configuration, and even per category.
\end{itemize}
From the above considerations, we infer an important finding: retaining only the diagonal terms and/or using a 
configuration truncated scheme will lead to unreliable results, as discussed in sec.~\ref{sec:disc}.

Finally, using values in the last line of Table~\ref{nsmm} our predictions for the proton strangeness magnetic moment 
$\mu_s$ are $-0.149 \pm 0.004 \mu_N$ for Set~I and $-0.067 \pm 0.004 \mu_N$ for Set~II, with the reported uncertainties
corresponding to the range $V = 570 \pm 46$ MeV~\cite{An:2011sb}. 

It is worth to underline two features: both Sets lead to small and negative values for $\mu_s$, though the two results 
differ one from another by more than $20\sigma$.
This latter observation shows the high sensitivity of the strangeness magnetic moment to the ratio $R=\omega_5/\omega_3$.
%
\begin{figure}[htb]
\begin{center}
\includegraphics[scale=0.7]{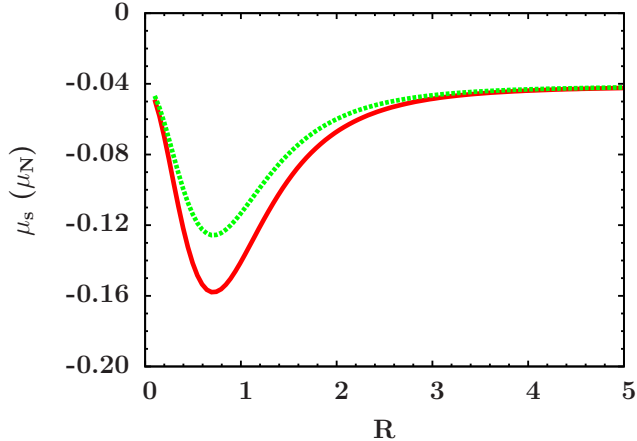}
\end{center}
\vspace{-1cm}
\caption{\footnotesize (Color online) The strangeness magnetic moment of the proton $\mu_s$ in units of nuclear magnetons ($\mu_N$) 
as a function of the ratio $R=\omega_{5}/\omega_{3}$, for $\omega_{3}=246$ MeV (full red curve) and $\omega_{3}=340$ MeV 
(dotted green curve).
\label{r35}}
\end{figure}

In Fig.~\ref{r35} $\mu_s$ is depicted as a function of $R$, varying from
$0.1$ to $5$, corresponding to the size of the strangeness component going from $10$ to $0.2$ times that of 
the three-quark configuration, with $\omega_{3}$ fixed at 246 MeV (full curve) and at 340 MeV (dotted curve).
The maximum discrepancy between the two curves is roughly 20\% at the minimum values for $\mu_s$,
located at $R \simeq 0.71$. 
So, $\mu_s$ depends mildly on the exact value of $\omega_{3}$, but strongly on that of $\omega_{5}$ and hence $R$.
The proton strangeness magnetic moment turns out then to be significantly sensitive to that ratio in the range
$0.1<R<3$, where $\mu_s$ varies by a factor of 4. 
In any case, according to our study, $\mu_s$ is small and negative.
%
%
\subsection{Strangeness magnetic form factor of the proton}
\label{sec:mff}
In order to extend the present approach to the $Q^{2}$-dependent strangeness magnetic form factor of the proton 
$G_{M}^{s}$, for which experimental data are available, we need to calculate the matrix elements of the transitions 
$\langle uuds\bar{s}|\vec{J}|uuds\bar{s}\rangle$ and $\langle uud|\vec{J}|uuds\bar{s}\rangle$ for both diagonal
and non-diagonal terms. 
For the former ones, explicit calculations lead to
\begin{equation}
\big (G_{M}^{s}\big )^{D}=\mu_{s}^{D} e^{-q^{2}/(5\omega_{5}^{2})}\,,
\label{g1}
\end{equation}
except for two of the configurations with four-quark subsystem wave functions being $[4]_{X}[31]_{FS}[211]_{F}[22]_{S}$ 
and $[4]_{X}[31]_{FS}[31]_{F}[22]_{S}$, for which the expression reads 
\begin{equation}
 \big (G_{M}^{s}\big )^{D}=\Big (\mu_{s}^{D}-\frac{2q^{2}}{15\omega_{5}^{2}}\Big )
 e^{-q^{2}/(5\omega_{5}^{2})}\,.
 \label{g2}
\end{equation}
For the non-diagonal transitions between all the strangeness configurations and the three-quark component of the proton,
the strangeness magnetic form factor is: 
\begin{equation}
\big ( G_{M}^{s} \big )^{ND}=\mu_{s}^{ND} e^{-4q^{2}/(15\omega_{5}^{2})}\,,
 \label{g3}
\end{equation}
with the photon three-momentum term ($q^{2}$)  related to the four-momentum transfer $Q=\sqrt{-k_{\gamma}^2}$ as
\begin{equation}
q^{2}=Q^{2}\Big (1+\frac{Q^{2}}{4M_{p}^{2}}\Big )\,. 
\end{equation}
\vspace{-0.5cm} 
{\squeezetable
\begin{table*}[hb]
\caption{\footnotesize Diagonal and non-diagonal contributions to the strangeness magnetic form factor of the proton from each
configuration for Sets I and II, at momentum transfer values $Q^2$=0.220 and 0.624 (GeV/c)$^2$.
\label{pmff}}
\scriptsize
\vspace{0.5cm}
\begin{tabular}{lllrrrrrrrrrrrrr} 
\hline\hline

                  & &&  \multicolumn{2}{l}{Set~I, $Q^2=0.220$}&&  \multicolumn{2}{l}{Set~II, $Q^2=0.220$} &&& 
                  \multicolumn{2}{l}{Set~I, $Q^2=0.624$}&&  \multicolumn{2}{l}{Set~II, $Q^2=0.624$}   \\
\hline

Category & $Configuration$   && $(G^{s}_{M})^D_i$ & $(G^{s}_{M})^{ND}_i$ && $(G^{s}_{M})^D_i$ & $(G^{s}_{M})^{ND}_i$ &&& 
$(G^{s}_{M})^D_i$ & $(G^{s}_{M})^{ND}_i$ && $(G^{s}_{M})^D_i$ & $(G^{s}_{M})^{ND}_i$  \\
\hline 
 i) $[31]_{X}[22]_{S}$: & & & & & & & && & & && & \\
 & $[31]_{X}[4]_{FS}[22]_{F}[22]_{S}$   && $.0039$  & $-.2918$   && $.0088$   & $-.2021$      &&& $.0005$   & $-.0206$   && $.0066$   & $-.1394$  \\
 & $[31]_{X}[31]_{FS}[211]_{F}[22]_{S}$ && $ .0016$ & $-.1778$   && $.0035$   & $-.1232$      &&& $.0002$   & $-.0126$   && $.0027$   & $-.0850$  \\
& $[31]_{X}[31]_{FS}[31]_{F}[22]_{S}$  && $ .0011$  & $-.1510$   && $.0025$   & $-.1046$      &&& $.0002$   & $-.0107$   && $.0019$   & $-.0721$  \\
 ~~~~{\it Subtotal 1} &                &&{\it .0066}&{\it --.6206}&&{\it .0148}&{\it --.4299}   &&&{\it .0009}&{\it --.0439}&&{\it .0112}&{\it --.2965}\\
%
%
 ii) $[31]_{X}[31]_{S}$: & & & & & & & && & & && & \\
& $[31]_{X}[4]_{FS}[31]_{F}[31]_{S}$   && $-.0051$   & $ .2334$   && $-.0112$   & $.1616$    &&& $-.0007$   & $.0165$   && $-.0085$   & $.1115$  \\

& $[31]_{X}[31]_{FS}[211]_{F}[31]_{S}$ && $-.0026$   & $ .1675$   && $-.0058$   & $.1160$    &&& $-.0004$    & $.0119$   && $-.0044$   & $.0801$  \\

& $[31]_{X}[31]_{FS}[22]_{F}[31]_{S}$  && $-.0014$   & $ .1241$   && $-.0032$   & $.0860$    &&& $-.0002$    & $.0088$   && $-.0024$   & $.0593$  \\
& $[31]_{X}[31]_{FS}[31]_{F}[31]_{S}$  && $-.0006$   & $ .0464$   && $-.0014$   & $.0325$    &&& $-.0001$    & $.0033$   && $-.0010$   & $.0225$  \\
 ~~~~{\it Subtotal 2} &                &&{\it --.0097}&{\it .5714}&&{\it --.0216}&{\it .3961}&&&{\it --.0014}&{\it .0405}&&{\it --.0163}&{\it .2734}\\
%
%
 iii) $[4]_{X}[22]_{S}$: & & & & & & & && & & && & \\
& $[4]_{X}[31]_{FS}[211]_{F}[22]_{S}$  && $-.0011$   & $ .2584$  && $-.0025$   & $.1790$   &&& $-.0002$   & $.0183$   && $-.0019$   & $.1235$  \\
& $[4]_{X}[31]_{FS}[31]_{F}[22]_{S}$   && $-.0009$   & $ .2258$  && $-.0019$   & $.1564$   &&& $-.0001$   & $.0160$   && $-.0015$   & $.1079$  \\
 ~~~~{\it Subtotal 3} &                &&{\it --.0020}&{\it .4842}&&{\it --.0044}&{\it .3354}&&&{\it --.0003}&{\it .0343}&&{\it --.0034}&{\it .2314}\\
%
%
 iv) $[4]_{X}[31]_{S}$: & & & & & & & && & & && & \\
& $[4]_{X}[31]_{FS}[211]_{F}[31]_{S}$  && $-.0062$   & $-.2455$   && $-.0138$   & $-.1700$   &&& $-.0009$   & $-.0174$   && $-.0104$   & $-.1173$  \\
& $[4]_{X}[31]_{FS}[22]_{F}[31]_{S}$   && $-.0035$   & $-.1511$   && $-.0078$   & $-.1047$   &&& $-.0005$   & $-.0107$   && $-.0059$   & $-.0722$  \\

& $[4]_{X}[31]_{FS}[31]_{F}[31]_{S}$   && $-.0015$   & $-.0705$   && $-.0034$   & $-.0488$   &&& $-.0002$    & $-.0050$   && $-.0026$   & $-.0337$  \\
 ~~~~{\it Subtotal 4} &                &&{\it --.0112}&{\it --.4671}&&{\it --.0250}&{\it --.3235}&&&{\it --.0016}&{\it --.0331}&&{\it --.0189}&{\it --.2232}\\ \\
& TOTAL                                && $-.0163$ & $-.0321$ && $-.0362$ & $-.0219$ &&& $-.0024$ & $-.0022$ && $-.0274$ & $-.0149$\\

\hline 
\hline
\end{tabular}
\end{table*}
}

\begin{figure}[ht]
\begin{center}
\includegraphics[scale=0.65]{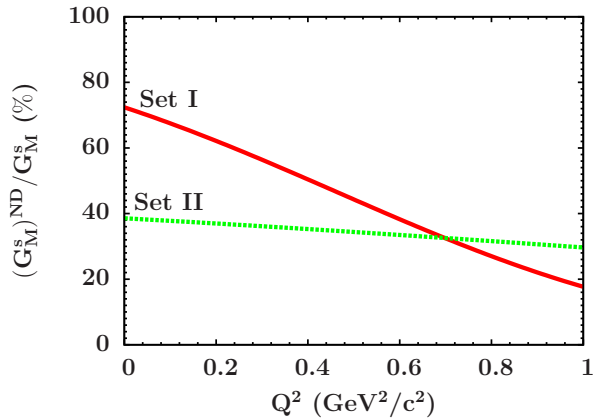}
\end{center}
\caption{\footnotesize (Color online) The ratio of non-diagonal to diagonal + non-diagonal terms in the strangeness magnetic form factor of 
the proton $G^s_M$ as a function of $Q^2$ for Sets I (full red curve) and II (dotted green curve).
\label{rndd}}
\end{figure}

Given the status of the data, discussed in the next section, we produce comprehensive numerical results at $Q^{2}=0.22$ and 
$0.624$~(GeV/c)$^2$.
Table~\ref{pmff} contains the outcome of our calculations on the proton strangeness magnetic form factor for all 12 configurations 
and for both Sets I and II, bringing in few comments:
\begin{itemize}
  \item {$\bf (G^{s}_{M})^D_i$  :} Because of the $\omega_5$ dependence of $G^{s}_{M}$, the diagonal terms are not 
  identical in Sets I and II, as it was the case for $\mu_s$. 
  The magnitude of this component, per configuration, decreases with $Q^{2}$ as well as in going from
  Set~II to Set~I at a fixed $Q^2$. 
  \item ${\bf (G^{s}_{M})^{ND}_i:}$ The magnitude of the non-diagonal terms are larger than those of diagonal ones, and they
  decrease with $Q^{2}$ and also in going from Set~I to Set~II at a fixed $Q^2$.
    \item {$\bf (G^{s}_{M})^D_i / (G^{s}_{M})^{ND}_i$:} the $Q^{2}$ dependence of this ratio turns out to be quite different
    for Sets I and II, as shown in Fig.~\ref{rndd}. For Set~I, between $Q^{2}=0$ and 1 (GeV/c)$^2$ the ratio decreases by a 
    factor of more than 3 and above $Q^{2}\sim0.4$ (GeV/c)$^2$, the diagonal terms become larger than the diagonal ones, while 
    in Set~II the non-diagonal terms stand for roughly $37 \pm 2\%$ of the sum of the two terms in the whole shown $Q^{2}$ range. 
    \item {\bf Signs:} There are no sign changes in diagonal and non-diagonal terms for a given configuration at different  
    $Q^{2}$s, including $Q^{2}=0$.
\end{itemize}
%
%
%

In the next section we proceed to comparisons between our results and relevant ones reported in the literature.
%
\subsection{Discussion}
\label{sec:disc}
\begin{figure}[hb]
\begin{center}
\includegraphics[scale=0.7]{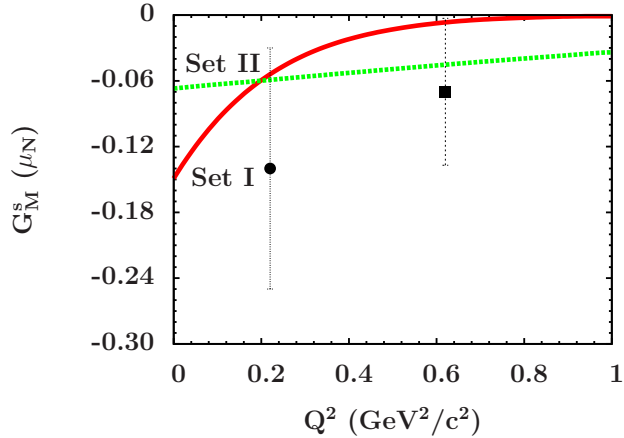}
\end{center}
\caption{\footnotesize (Color online) The strangeness magnetic moment of the proton $G^s_M$ as a function of the 
momentum transfer $Q^2$ for Sets I (full red curve) and II (dotted green curve). 
Data are from Refs.~\cite{Baunack:2009gy,Ahmed:2011vp}.
\label{mmq}}
\end{figure}

Table~\ref{results} summarizes our numerical results for the strangeness magnetic moment of the proton and its magnetic  
form factor at four $Q^{2}$ values.
In Fig.~\ref{mmq} results for $G_M^s$ within Sets I and II, spanning the range $0 \leq Q^{2} \leq 1$ (GeV/c)$^2$ are depicted and 
compared to the HAPPEX~\cite{Ahmed:2011vp} and PVA4~\cite{Baunack:2009gy} data.
%
%
\begin{table}[ht]
\caption{\footnotesize Results for the proton strangeness magnetic moment and magnetic form factor (in nuclear magneton) at four $Q^2$ values
(in (GeV/c)$^2$). 
\label{results}}
\scriptsize
\vspace{0.5cm}
\begin{tabular}{lcccccc} 
\hline\hline
        Reference~{Year}    & Approach  & $\mu_s$   &  $G^{s}_{M}(Q^2=0.10)$     &  $G^{s}_{M}(Q^2=0.22)$     & $G^{s}_{M}(Q^2=0.62)$ &  $G^{s}_{M}(Q^2=0.81)$ \\ 
\hline
{\bf Present work: Set~I}  &$E \chi CQM$& $-0.149 \pm 0.004$ &  $-0.093 \pm 0.002$ & $-0.051 \pm 0.004$ & $-0.006\pm 0.000$  & $-0.002\pm 0.000$     \\
{\bf Present work: Set~II} &            & $-0.067 \pm 0.004$ &  $-0.063 \pm 0.004$ &$-0.059 \pm 0.004$ & $-0.045 \pm 0.003$  & $-0.039 \pm 0.003$    \\
%
%
Leinweber~{\it et al.}~\cite{Leinweber:2004tc}~$\{2005\}$& LQCD & $-0.046 \pm 0.019$ &  &&                            &  \\
Wang~{\it et al.}~\cite{Wang:1900ta}~$\{2009\}$          & LQCD &  &  & $-0.034 \pm 0.021$&                            &  \\
Doi~{\it et al.}~\cite{Doi:2009sq}~$\{2009\}$            & LQCD & $-0.017 \pm 0.026$ & $-0.015 \pm 0.023$ &           &&  \\
Babich~{\it et al.}~\cite{Babich:2010at}~$\{2012\}$      & LQCD &                    &  &   $-0.002 \pm 0.011$       & $-0.007\pm 0.012$ & $-0.022\pm 0.016$ \\
Ahmed {\it et al.}~\cite{Ahmed:2011vp}~$\{2012\}$    &Data [HAPPEX]  & &                   &                             & $-0.070 \pm 0.067$ &     \\
Baunack {\it et al.}~\cite{Baunack:2009gy}~$\{2009\}$& Data [PVA4] &   &                &$-0.14 \pm 0.15$    &  &                    \\
Androic {\it et al.}~\cite{Androic:2009aa}$\{2010\}$  & Data [G0]   & &                   &$+0.083 \pm 0.217$  & $-0.123 \pm 0.130$ & \\
Spayde {\it et al.}~\cite{Spayde:2003nr}$\{2004\}$  & Data [SAMPLE]   & & $+0.37 \pm 0.34$ &  &  & \\
%
%
%
\hline
\hline
\end{tabular}
\end{table}
%
%

The general trend in our results is that the investigated observable is negative with small magnitude.
However, Sets~I and II behave differently as a function of $Q^2$. 
Actually, for Set~I, the harmonic oscillator parameter $\omega_{5}\simeq225$~MeV, is smaller than  
$\omega_{5}\simeq600$~MeV in Set~II. 
So due to the exponential $Q^{2}$ dependence, $G^{s}_{M}$ approaches zero faster in Set~I than in Set~II.
In the following we compare our predictions with results from other sources quoted in Table~\ref{results}. 

At $Q^2$=0.22 (GeV/c)$^2$ both Sets give almost identical values, compatible with PVA4 data~\cite{Baunack:2009gy},
while at $Q^2$=0.624 (GeV/c)$^2$ Set~II is favored by the HAPPEX~\cite{Ahmed:2011vp} data. 
At those two momentum transfer values, data reported by the G0 Collaboration~\cite{Androic:2009aa} have too 
large uncertainties to allow informative comparisons with our predictions.
To a lesser extent, the same consideration is also true for the SAMPLE Collaboration data~\cite{Spayde:2003nr} 
at $Q^2$=0.1 (GeV/c)$^2$ with a positive value, large uncertainty and compatible with zero.

In Table~\ref{results} we also show results from lattice-QCD calculations. 
Quenched QCD complemented by chiral extrapolation techniques performed by Leinweber~{\it et al.}~\cite{Leinweber:2004tc} 
and Wang~{\it et al.}~\cite{Wang:1900ta} produce for $\mu_s$ and $G^{s}_{M}(Q^2=0.22)$, respectively, theoretical data
compatible with our predictions within less than $2\sigma$ for $\mu_s$ in Set II and $G^{s}_{M}$ in both Sets. 
This is also the case for Set II results with respect to the outcome of a $N_f=2+1$ clover fermion LQCD by 
Doi~{\it et al.}~\cite{Doi:2009sq} for $\mu_s$ and $G^{s}_{M}(Q^2=0.10)$, albeit with large uncertainties and smaller, 
central values in magnitude.
Finally, a recent exploratory calculation by Babich~{\it et al.}~\cite{Babich:2010at}, based on the Wilson gauge and 
fermion actions on an anisotropic lattice, leads to smaller magnitudes than our predictions at $Q^2=0.22$ (GeV/c)$^2$. 
While at $Q^2=0.62$ (GeV/c)$^2$ result of the latter work agrees with ours for Set~I, at $Q^2=0.81$ GeV/c)$^2$ Set II 
produces value compatible with the considered LQCD data.

Here, it is worth mentioning that theoretical predictions as well as recent data (Table~\ref{results}) show (significant) 
discrepancies with the extracted values from  global fits to the data released before 2009: 
$G^s_M (Q^2=0.22) = 0.12 \pm 0.55 \mu_N$ (Ref.~\cite{Young:2006jc}), $G^s_M (Q^2=0.21) = 0.19 \pm 0.21 \mu_N$. 
(Ref.~\cite{Liu:2007yi}) and $G^s_M (Q^2=0.624) = 0.08 \pm 0.11 \mu_N$ (Ref.~\cite{Pate:2008va}), all of them in 
disagreement with the latest data from PVA4~\cite{Baunack:2009gy} and HAPPEX~\cite{Ahmed:2011vp} Collaborations. 

To end this section, we compare our approach to results coming from similar 
works~\cite{Zou:2005xy,Riska:2005bh,An:2006zf,An:2005cj,Kiswandhi:2011ce} reported in the literature.

As mentioned in Introduction, in Ref.~\cite{Zou:2005xy} the sign of the proton strangeness magnetic moment was 
investigated with respect to the strange antiquark states in the five-quark component of the proton. 
In a subsequent paper~\cite{Riska:2005bh} the authors calculated $G_M^s (Q^2)$ in the range 
$0 \leq (Q^2)\leq 1$ (GeV/c)$^2$, where data were giving positive values~\cite{Young:2006jc,Liu:2007yi,Pate:2008va}.
There, two scenarios were adopted i) $\omega_5 \simeq 2 \omega_3$ ($R \simeq$2) and 
$\omega_5 \simeq \omega_3$ ($R \simeq$1), and also two values for the probability of the $s \bar s$, namely 
$P_{s \bar s}$= 10\% and 15\%.
The three combinations between $R$ and $P_{s \bar s}$ studied gave results consistent with the available data in 2006.
However, out of the twelve configurations (Table~\ref{nsmm}) only $[31]_{X}[4]_{FS}[22]_{F}[22]_{S}$ was considered.
That configuration was also used in Refs.~\cite{An:2006zf,An:2005cj}, where only the diagonal term was included, resulting
in $\mu_s$=0.17$\mu_N$.

A more recent constituent quark model~\cite{Kiswandhi:2011ce} considered separately  only two configurations, namely, 
$[31]_{X}[4]_{FS}[22]_{F}[22]_{S}$ and $[31]_{X}[31]_{FS}[211]_{F}[22]_{S}$, corresponding to the $\bar s$ being in the $S-$ 
or $P-$state, respectively.
Pure $P$-state gave $\mu_s=0.066\mu_N$ and an admixture between the two states $\mu_s=1.01 \mu_N$. 
In that work, both diagonal and non-diagonal terms were considered  for the retained configurations and $\omega_3$ 
was fixed at 246 MeV, while $\omega_5$ and the probability $P_{s \bar s}$ were fitted on the G0 
Collaboration~\cite{Androic:2009aa} data reported in Table~\ref{results}.
The extracted values are $\omega_5$=469 MeV and $P_{s \bar s}$=0.025\%, smaller by more than two orders of magnitude compared 
to the $^3P_0$ model result employed in the present work. 
Using their approach, the authors found that putting $P_{s \bar s}$=2.5\%, as reported in 
Ref.~\cite{arXiv:1102.5631}, leads to $\omega_5$=108 MeV.
The incredibly tiny probability reported in Ref.~\cite{Kiswandhi:2011ce} can easily be understood. 
As shown in Table~\ref{pmff}, contributions from individual configurations 
$[31]_{X}[4]_{FS}[22]_{F}[22]_{S}$ or $[31]_{X}[31]_{FS}[211]_{F}[22]_{S}$
compared to the total of contributions from all twelve of them differ by up to two orders of magnitude.
%
%
\section{Summary and conclusions}
\label{sec:end}
The extended chiral constituent quark model offers an appropriate frame to study the possible manifestations of
genuine five-quark components in baryons. The present work is in line with our earlier 
efforts~\cite{An:2010wb,An:2011sb,An:2012kj} in that realm.
There are several difficulties in this endeavor: few observables have been identified carrying information
on higher Fock states, the data are scarce and often bear large uncertainties due to the smallness of the
effects looked for. Moreover, there are input parameters in the approach, which basically should be taken
from literature and exceptionally fitted on the data under consideration. 
Accordingly, we took advantage of the data on radiative and strong decays of the $\Lambda(1405)$ resonance~\cite{An:2010wb},
strong decay of low-lying $S_{11}$ and $D_{13}$ nucleon resonances~\cite{An:2011sb}, and
sea flavor content of octet baryons~\cite{An:2012kj} to deepen our understanding of the five-quark components
and select a coherent set of input parameters.

Our main findings can be summarized in three points, as follows.
\begin{itemize}
  \item {\it i)} {\bf Five-quark Fock states:} we gave detailed numerical results for both diagonal and non-diagonal 
  terms for all of the twelve relevant configurations showing strong interplays among different components 
  with (very) large cancellations.
  \item {\it ii)} {\bf Probability of the $s \bar s$ in the proton wave function:} we determined  $P_{s \bar s}$ using a $^{3}P_{0}$ 
pair creation model, as in a previous work~\cite{An:2012kj}.
  \item {\it iii)} {\bf Harmonic oscillator parameters:} it was shown that with respect to the parameters $\omega_3$ and $\omega_5$,
the important element is the ratio $R=\omega_5/\omega_3$.
\end{itemize}
Based on the above observations, it becomes then obvious that using severely truncated configuration sets
and/or unrealistic values for $P_{s \bar s}$ or $R$ will lead to unreliable results with respect to the magnetic 
moment and/or magnetic form factor of the proton.

In the present paper we showed that our predictions are in reasonable agreement with recent measurements
~\cite{Baunack:2009gy,Ahmed:2011vp} and lattice-QCD results~\cite{Leinweber:2004tc,Wang:1900ta,Doi:2009sq,Babich:2010at}.

The uncertainties associated to the available data on the one hand, and those of LQCD approaches on the other hand, 
do not allow us making a sharp choice between the results coming from the two Sets in terms of the ratio $R$.
It is nevertheless clear that the strangeness magnetic moment of the proton and its magnetic form factor 
are small and negative. 
Between the two Sets, Set~II appears to be slightly favored by findings from other sources. 
Accordingly, we get $\mu_s = -0.0670 \pm 0.004 \mu_N$ and the magnitude of the strangeness magnetic from factor 
of the proton evolves smoothly with increasing transfer momentum to reach $G_M^s(Q^2) = -0.033 \pm 0.003 \mu_N$ at $Q^2=1$ (GeV/c)$^2$.

Awaited for data at $Q^2=0.6$ (GeV/c)$^2$ expected to be released by the PVA4 Collaboration ~\cite{Baunack:2009gy}
and more advanced LQCD approaches will hopefully improve the accuracy of the experimental and theoretical data bases. 
Recent convergence between theory and experiment on the negative sign of that observable and its smallness, might also
initiate new dedicated measurements.
\begin{acknowledgments}
We wish to thank the anonymous Referee for his/her careful reading of the manuscript. 
This work was supported by the National Natural Science Foundation of China under grant number 11205164.
\end{acknowledgments}
%
%
%

%
%
\end{document}